\documentclass[twocolumn,final,showpacs,preprintnumbers,amsmath,amssymb]{revtex4}
 \usepackage[dvips]{graphicx}
 \usepackage{color}
\begin{document}

\preprint{}

\title{The calculation of multi-fractal properties of directed random walks on hierarchic trees with continuous branching }
\author{David B. Saakian$^{1,2,3}$}
\email{saakian@yerphi.am}

 \affiliation{$^1$Institute of
Physics, Academia Sinica, Nankang, Taipei 11529, Taiwan}
\affiliation{$^2$ A.I. Alikhanyan National Science Laboratory
(Yerevan Physics Institute) Foundation, 2 Alikhanian Brothers St.,
Yerevan 375036, Armenia}

\affiliation{$^3$National Center for Theoretical Sciences (North):
Physics Division, National Taiwan University, Taipei 10617, Taiwan}
\begin{abstract}
We consider the hierarchic tree Random Energy Model with continuous branching and
calculate the moments of the corresponding partition function. We establish the
multifractal properties of those moments. We derive formulas for the normal
distribution of random variables, as well as for the general case.
We compare our results for the moments of partition function with
corresponding results of logarithmic 1-d REM and conjecture a specific powerlaw tail for the partition function distribution in the
high-temperature phase. Our results establish
a connection between reaction-diffusion equations and multi-scaling.
 \end{abstract}

\pacs{11.25.Mj,75.10.Nr} \maketitle

\maketitle
\section{ Introduction}

Random energy models on hierarchic trees are rather well-known and much investigated
objects of statistical physics
\cite{de88},\cite{de90},\cite{de91},\cite{de93}. The
tree is constructed via a deterministic branching process, starting from the root node
and adding $q$ branches from every node of the tree, so that after K steps of the procedure there are $q^K$ nodes
of the last generation (the endpoints of the tree).
One then associates random energy variables to every branch of the tree, and further attributes the
random energy to every node of the last generation  by adding up all the energy variables along the unique
path connecting the given endpoint to the root. At the last step of the procedure
one builds up the partition function \cite{de88}-\cite{de91}.
 In the context of statistical physics the model has been identified with directed polymers on the trees,
 and has been also related to the spin-glass model REM
\cite{de80},\cite{de89}. Independently, similar models have been applied and
extensively investigated also in the context of financial mathematics and
turbulence \cite{ma97},\cite{ma97a},\cite{ba06},\cite{st07},\cite{tu}. They are also intimately connected
to 2-d conformal models \cite{ch96},\cite{ca97}. The hierarchical structure of the tree naturally induces recursive
relations for the partition function, with the branching number $q$ being a parameter in the recursion.
 While by construction the branching number q is an
integer, one can formally consider the recursive equations
\cite{ba},\cite{de88} in the $q\to 1$ limit, simultaneously allowing $K\gg 1$ and keeping
the number of endpoints $q^K$ fixed. Such an idea has been suggested in
\cite{sa02}, later worked out in more detail in \cite{sa09}. Along that line
an exact renormalization equation for continuous tree models has been derived in \cite{sa12}
for the general distribution of random variables.

The hierarchic tree models belong to the type of  Random Energy
Models (REM's) which all share the leading term in the free energy
 with that of the simplest Derrida's model. The investigation of REM like models in finite dimensions has been started in \cite{ca01}
and got serious impetus from recent solution of 1-d models with logarithmic correlation
of energies at different sites \cite{bu08,fy09}. That solution essentially used the generalization of Selberg integrals
\cite{se44},\cite{se08} to complex number of integrations (see rigorous mathematical justification in \cite{Ost}).
 The 1d models are intimately connected with the model \cite{ba00},\cite{ba01}. Recently \cite{sa12a}
we have developed a statistical physics approach to related dynamical MSM
models \cite{cf02}. While all three types of REM models:  hierarchical
tree \cite{de88}, 1-d logarithmic REM \cite{fy09} and MSM
\cite{cf02} have exactly the same mean free energy as the standard REM, they have different
distributions for the free energy and the partition function. Most essentially, in the Derrida's  REM the probability
distribution of the partition function has no fat tails in the high temperature
phase \cite{de89}, while such tails are present in 1d models \cite{bu08,fy09}.
In this work we give indication that such tails exists in the hierarchic model, and also reveal
 the multi-fractal properties of the later, which is important for applications
\cite{ma97}-\cite{st07}. For a recent discussion of multifractal
properties of REM, and other models with logarithmic correlations
see \cite{fy09a,fy10} and references therein.

In multifractal approach \cite{je86},\cite{pa87} one considers the
moments of a random variable (partition function) Z at some scale
$l$ defined with respect to the maximal scale L:
\begin{eqnarray}
\label{e1} <Z^n>= e^{\xi(n)\ln \frac{l}{L} }\hat Z_n,
\end{eqnarray}
where the exponent $\xi(n)$ defines the multi-scaling.

Knowing the $\xi(n)$ and the coefficients $\hat Z_n\sim  O(1)$ in
the limit of large $L$, one can reconstruct the probability
distribution of the partition function. In this way such a distribution has been explicitly derived
 for 1-d logarithmic REMs \cite{bu08,fy09}

In the present article we will calculate $\xi(n)$ for hierarchical
tree, and derive recursive relations to define $\hat Z_n$, which, in
principle, could be applied to calculate the probability
distribution of Z within some accuracy. We shall see indications of
the fact that the divergence of moments is essentially the same as
for $1d$ case, thus the two types of models must share the same fat
tails.

\section{The calculation of the moments }

\subsection{The model with Normal distribution of random variables}

Let us define the model outlined in the introduction following the papers
\cite{sa02},\cite{sa12}. Starting with the hierarchical tree with
integer branching number  $q$ and $q^K$ endpoints, we consider two such end-points $w_i$ and $w_j$.
The two paths connecting these points to the root coincide up to a level m, counting from the root.
Accordingly, we can define the hierarchic
distance between the two points as
\begin{eqnarray}
\label{e2} v(w_i,w_j)=\frac{mV_0}{K}
\end{eqnarray}
where $V_0$ is defined as
\begin{eqnarray}
\label{e3} q^K\equiv e^{V_0}\equiv L
\end{eqnarray}
Associated with every branch of the tree is a Gaussian random
variable $\epsilon$ distributed according to the law
\begin{eqnarray}
\label{e4}
\rho(\frac{V_0}{K},\epsilon)=\sqrt{\frac{K}{2V_0\pi}}\exp[-\frac{K\epsilon^2}{2V_0}]
\end{eqnarray}
We define the energy $y_i$ at the endpoint $w_i$ as a sum of
corresponding variables $\epsilon$ sampled along the unique path connecting the
point $w_i$ to the of the tree.

We further define the partition function
\begin{eqnarray}
\label{e5} Z= \sum_{i}e^{-\beta y_i-V_0\frac{\beta^2}{2}},
\end{eqnarray}
where the sum is over $e^{ V}\equiv l$ end-points.

Then we obtain
\begin{eqnarray}
\label{e6} <Z(e^{ V},b)^n>=  \sum_{i_1,...i_n}< e^{-\beta
y_{i_1}...-\beta y_{i_n}}>
\end{eqnarray}
In Eq.(\ref{e6}) the sum is over all endpoints of the tree separated
by the hierarchic distance $ V$ from a chosen point.

Now we consider the limit
\begin{eqnarray}
\label{e7} q\to 1
\end{eqnarray}
while keeping the value of $V_0$ fixed and large. Recall that $V_0=K
\log q$. In such a limit there are $e^v$ end points at the
hierarchic distance $v$. While calculating the $<Z^n>$, we may
discard the contributions  given by Fig.~1b.

We formally replace the sum over the tree  with the integration over
a measure $d w$ \cite{sa02,sa09},
\begin{eqnarray}
\label{e8} <Z(e^{ V},b)^n>=
\prod_{i=1}^n \int dw_i<e^{-\beta \sum_iy(w_i)}>
\end{eqnarray}
We are able to calculate $<Z^n>$ for the integration range going
over the maximal distance $ V$.
\begin{figure} \large \unitlength=0.1in
\begin{picture}(42,12)
\put(-1.7,-2){\includegraphics{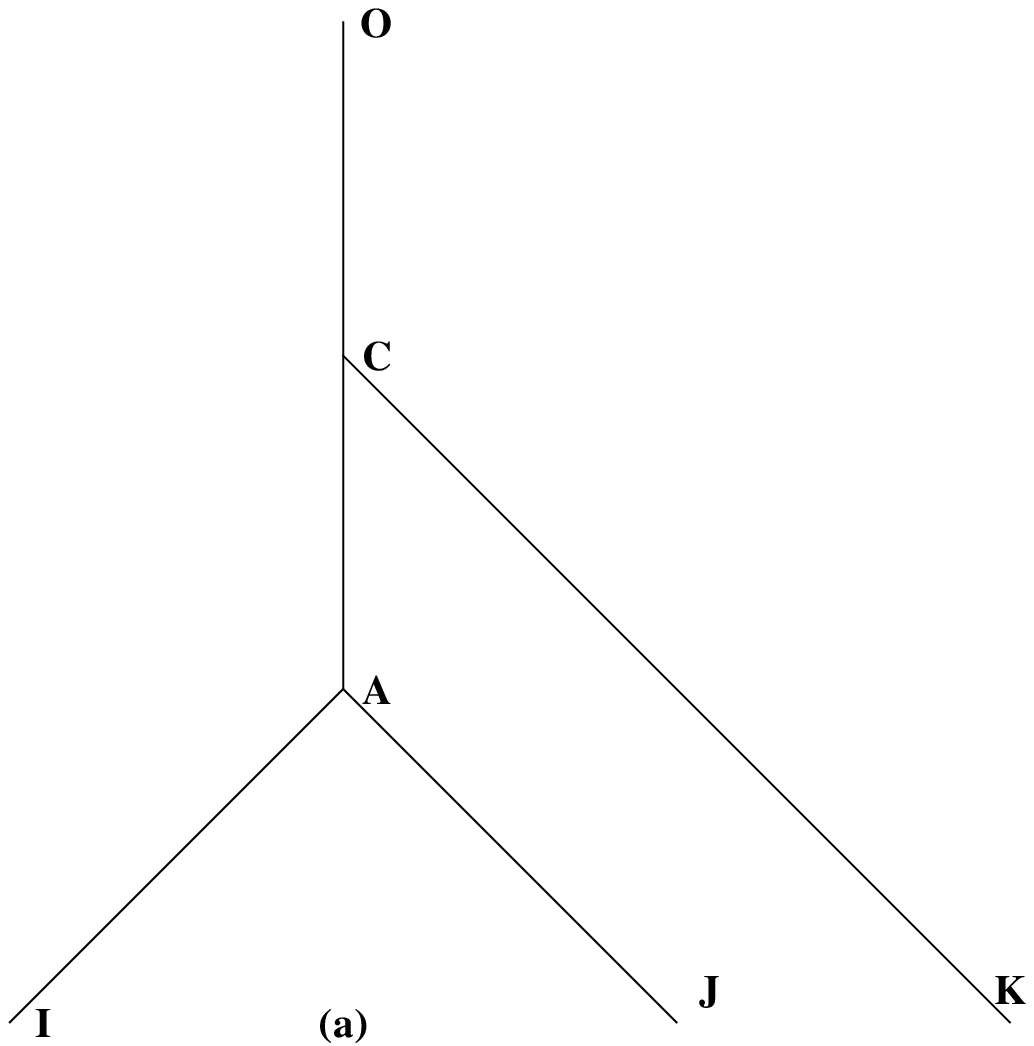}}
\put(16.5,-2){\includegraphics{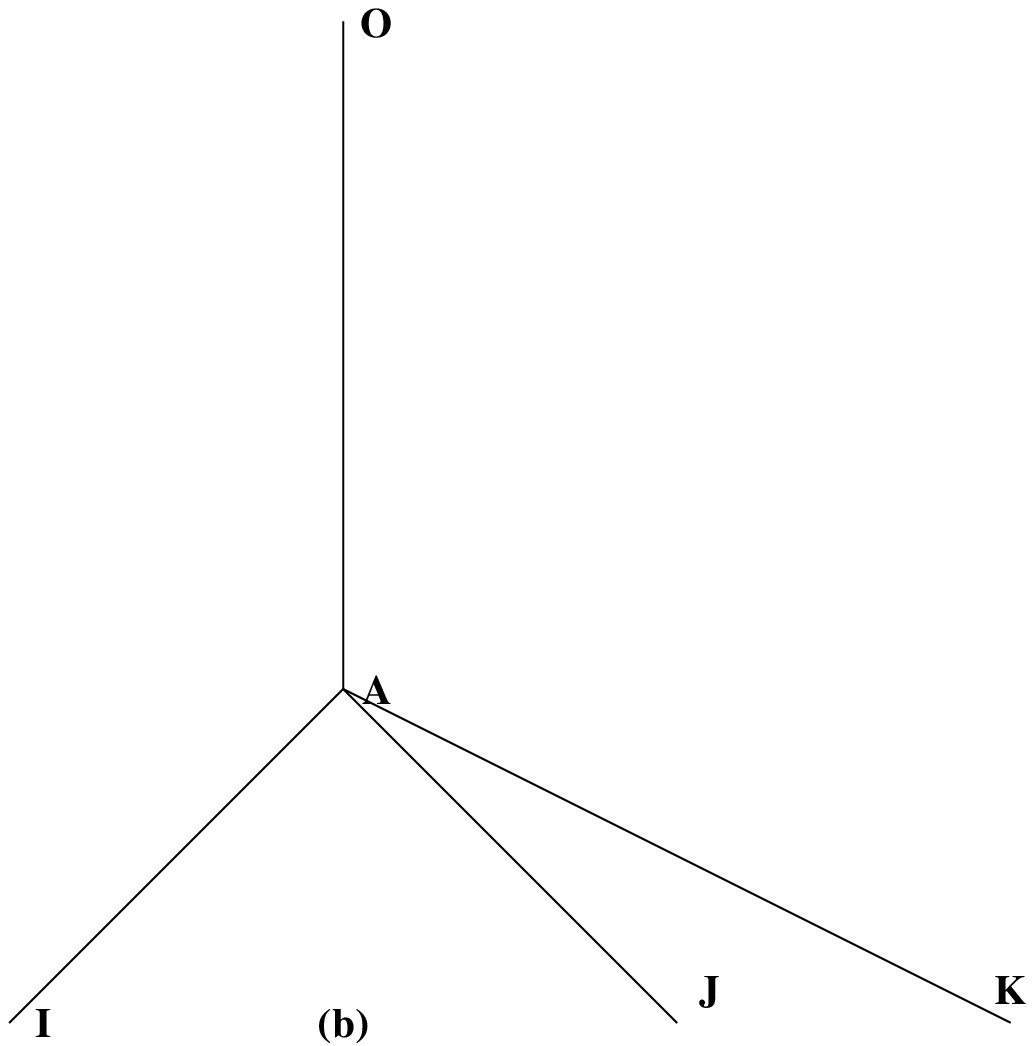}}
\end{picture}
\caption{The sum is over three marked endpoints $I,J,K$ of the
hierarchic tree. Only the Fig. 1.a contributes to the leading
order. The locations of $I,J,K$ given by Fig. 1.b yield negligible
$O(q-1)$ contribution to the partition function in Eq.(8).}
\end{figure}
When calculating the correlations in Eq.(8) we use the following
trick: the part of the trajectory of $w_i$ with the hierarchic
length $v$ which has no overlap with trajectories from other points
yields a factor $e^{\beta^2v/2}$, while the part  of the trajectory
common to $n$ such trajectories and hierarchic length $v$ yields a
factor $e^{n^2\beta^2v/2}$.

Consider $n=2$ for simplicity, with summation going up to the maximal distance $V$.
We then have $e^{V-v}$ positions for the top level of hierarchy. Then, summing (or "integrating")
 over the number of two points at the distance $e^{2v}$ we get
\begin{eqnarray}
\label{e9}
<Z(e^V,b)^2>=\int_0^{V}dve^{V+v}e^{2(V_0-v)\beta^2+v\beta^2-V_0\beta^2}\nonumber\\
=\frac{1}{1-\beta^2}e^{V(2-\beta^2)+V_0\beta^2}=\nonumber\\
=\frac{1}{1+b}e^{V(2+b)-bV_0}
\end{eqnarray}
where we have introduced the parameter $b=-\beta^2$.

Consider now the integration in Eq.(\ref{e8}) going up to the maximal hierarchic distance
V. We assume the following Ansatz:
\begin{eqnarray}
\label{e10}<Z(e^V,b)^n>\equiv e^{(V-V_0)(n+bn(n-1)/2)}e^{nV_0}\hat
Z_n
\end{eqnarray}
where we have introduced dimensionless  coefficients $\hat Z_n$.
The results of our calculation support the ansatz in Eq.
(\ref{e10}), see Eqs.(\ref{e13})-(\ref{e14}).

Actually the ansatz Eq.(\ref{e10}) is correct only when
\begin{eqnarray}
\label{e11} (n-1)-n\frac{b}{2}>-\frac{n^2b}{2},
\end{eqnarray}
otherwise there is a phase transition to a different phase \cite{de89},
which manifests itself via diverging integration in Eq.(\ref{e8}).

We will explicitly consider the case of positive $b$ (imaginary
$\beta$) while calculating some eventual expression (there is no any
restriction on $n$ like Eq.(11) ), like the analytical approximation
of the fractional moment
\begin{eqnarray}
\label{e12} <Z^{\alpha}( V,b)>,
\end{eqnarray}
and then continue analytically the resulting expressions to the
realistic case of negative $b<0$. The analytical continuation
gives wrong results in SG, when we continue the expressions for
moments to the other statistical physics phase. Hopefully in our
case we are interested to continue to expressions of moments in
the same  phase.

Considering now the general case we calculate $Z_n$ recursively.  At
the highest hierarchy level,  $n$ points are split into two
groups with $m$ and $(n-m)$ points,$1\le m<n$ accordingly, see Fig.2. Having
the expressions of $Z_m,Z_{n-m}$ at our disposal, we can calculate $Z_n$.
 We assume the
minimum hierarchy level where all the paths from the root to $n$ points meet is given by $v$. There
are $e^{V-v}$ such locations on the tree. Integrating over the
positions of all points with a given {\rm v} we arrive to
\begin{eqnarray}
\label{e13} Z^n(e^V,b)=\sum_{1\le m<n}\times\nonumber\\
\int_0^V dv
[Z_m(v)Z_{n-m}(v)]e^{V-v+bvm(n-m)}
\end{eqnarray}
 Using the scaling Ansatz Eq.(\ref{e10}) gives
 \[
  \hat Z_{n}(b)=\sum_{1\le m<n}\times
 \]
\[
\frac{ \hat Z_m(b)\hat Z_{n-m}}{n-1+\frac{b}{2}
(m(m-1)+(n-m)(n-m-1)-2m(n-m))}
\]
\begin{eqnarray}
\label{e14}
=\frac{\sum_{1\le m<n} \hat Z_m(b)\hat
Z_{n-m}(b)}{(n-1)(1+nb/2)}
\end{eqnarray}
 Thus, the problem amounts to solving the recursive equation (\ref{e14}) with the initial
condition (\ref{e12}).

Eq.(\ref{e10}) defines the so-called multifractal scaling. We can rewrite it in the form
\begin{eqnarray}
\label{e15}<Z(l,b)^n>\equiv e^{(n+bn(n-1)/2)\ln (l/L)}\hat Z_n
\end{eqnarray}

Let us now define a characteristic function
\begin{eqnarray}
\label{e16} u(x)=\sum_{n=1}\hat Z_nx^{n}
\end{eqnarray}
for those values of $\beta$ where the sum converges. We rewrite
Eq.(14) as
\begin{eqnarray}
\label{e17}(n-1)(1+nb/2)\hat Z_n(b) =\sum_{1\le m<n} \hat Z_m(b)\hat
Z_{n-m}(b)
\end{eqnarray}
Using the equations:
\begin{eqnarray}
\label{e18}\sum_nx^n\sum_{1\le m<n} \hat Z_m(b)\hat
Z_{n-m}(b)=(\sum_nx^n \hat Z_n(b))^2,\nonumber\\
\sum_n n\hat Z_n(b)x^n=x\frac{d}{dx}\sum_n \hat Z_n(b)x^n,\nonumber\\
\sum_n n^l\hat Z_n(b)x^n=(x\frac{d}{dx})^l\sum_n \hat Z_n(b)x^n,
\end{eqnarray}
we derive the following ODE for the characteristic function:
\begin{eqnarray}
\label{e19}
\frac{bx^2}{2}\frac{d^2u(x)}{dx^2}+x\frac{du}{dx}=u(x)+u^2(x),\nonumber\\
u(0)=1,u'(0)=1
\end{eqnarray}
Unfortunately we could not solve Eq.(19)  to derive explicitly
analytical expressions for $Z_n$ in case of general (complex) values
of $n$, which precludes from following the procedure of
\cite{bu08,fy09}. Nevertheless, Eq.(19) allows us to extract a few
first moments.  We find:
\begin{eqnarray}
\label{e20}\hat Z_1=1,\hat Z_2=\frac{1}{1+b}; \hat Z_3=\frac{2}{(1+b)(2+3b)}\nonumber\\
\hat Z_4=\frac{4}{(1+b)(2+3b)(3+6b)}+\frac{1}{(1+b)^2(3+6b)}
\end{eqnarray}
\begin{figure} \large \unitlength=0.1in
\begin{picture}(42,12)
\put(0,0){\includegraphics{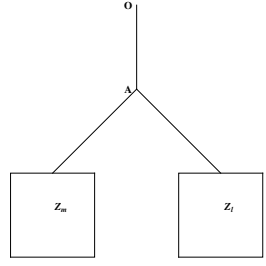}}
\end{picture}
\caption{The graphics for calculation of $Z_{n}$ recursively
fracturing  $ n=l+m$. The groups with $ l$ and $ m$ endpoints meet
at the hierarchic level v at the point A. }
\end{figure}

The above relation allows us also to calculate the asymptotic of $\hat Z_n$ at large $n$.
In doing this we assume that our function $u(x)$ has a singularity at some $\bar
\rho$ of the form \cite{fr}
\begin{eqnarray}
\label{e21} u(x)=\frac{c}{(1-\frac{z}{\bar\rho})^{\alpha}}
\end{eqnarray}
Putting the latter expression into Eq.(\ref{e19}), we derive:
\begin{eqnarray}
\label{e22} \alpha=2,\nonumber\\ c=\frac{3b^2}{\bar\rho^2}
\end{eqnarray}
Eq.(\ref{e22}) gives the asymptotic expression
\begin{eqnarray}
\label{e23} \hat Z_n=\frac{3(n+1)b^2}{\bar\rho^{2+n}}
\end{eqnarray}
We can define the value of $\bar\rho$ only numerically, looking at
$Z_n$ for large values of the parameter $n$. A similar problem has
been considered in \cite{hw10}. Slightly modifying their results we
arrive at the following algorithm to calculate $\bar\rho$. For a
given integer $n_0$ we are looking the minimum over $j$ and obtain:
\begin{eqnarray}
\label{e24}
\bar\rho=min[\frac{6(n-1+n(n-1)b^2/2)}{n(n+2)p_j}]^{1/(2+n)} |_{0\le
j\le n_0}
\end{eqnarray}
It is possible to get accurate values of $\bar \rho$ by increasing the
value of $n_0$.

In \cite{sa02},\cite{sa12} the
following reaction-diffusion equation has been derived for the same model:
\begin{eqnarray}
\label{e25} \frac{\partial G(x,v)}{\partial v}=\frac{\partial^2
G(x,v)}{2\partial x^2}+G(x,v)\ln G(x,v)
\end{eqnarray}
where $0\le v\le V$ played the role of time parameter in the
reaction-diffusion equation. We should solve that equation with the
initial value of $G$
\begin{eqnarray}
\label{e26} G(x,0)=\exp[-e^{-\beta x}]
\end{eqnarray}
Therefore, there must be a relation between ODE (\ref{e19}) and PDE
(\ref{e25}): solving Eq.(\ref{e25}) one can find the solutions of
Eq.(\ref{e19}). Eq.(\ref{e25}) allows one to be investigated via the
traveling wave approach \cite{sa12} and $\beta^2=2$ is its critical
point. The critical point of Eq.(19) should therefore correspond to
$b=-2$, and is associated with the anticipated freezing transition
from the high temperature phase to spin-glass-like phase.

\subsection{The case of general distribution}

Instead of Eq.(4)  we now consider
\begin{eqnarray}
\label{e27}
\rho(\frac{V_0}{K},x)=\frac{1}{2\pi}\;\int_{-\infty}^{\infty}\; dh
\;\exp[\frac{V_0}{K}\phi(ih)-ihx]
\end{eqnarray}
Then for the the sum $y_i$ of K random variables x, we can compose
the distribution $\rho(V_0,\epsilon)$ simply by multiplying $\phi$
in the exponent of Eq. (27) by $K$:
\begin{eqnarray}
\label{e28}
\rho(V_0,\epsilon)=\frac{1}{2\pi}\;\int_{-\infty}^{\infty}\; dh
\;\exp[V_0\phi(ih)-ih\epsilon]
\end{eqnarray}
We define
\begin{eqnarray}
\label{e29} Z= \sum_{i}e^{-\beta y_i-V_0\phi(\beta)},
\end{eqnarray}
We are  calculating the correlations in $<Z^n>$ using the following
trick: the part $v$ of the trajectory of $w_i$, which belongs only
him (no intersection with trajectories of other points), gives a
factor $e^{\phi(\beta)v}$, while the part v of trajectory common
to $n$ trajectories, gives the factor $e^{\phi(n\beta)v}$.

Repeating the calculations of the previous section we obtain:
\begin{eqnarray}
\label{e30} <Z^2>=\int_{0}^Vdv
e^{(V-v)\phi(2\beta)}e^{2v\phi(\beta)-2V_0\phi(\beta)}e^{2v}e^{V-v}
\end{eqnarray}
The factor $e^{(V-v)\phi(2\beta)}$ corresponds to the part of
trajectories  $[A,O]$, see Fig1.a, while any of the lines $[I,A]$
and $[J,A]$ gives the factor $e^{2v\phi(\beta)}$, with $e^{2v}$ being the
number of possible locations of $I,J$ at the hierarchic distance
$v$, whereas $e^{V-v}$ is the number of possible positions of the point
$A$.

Performing the integration over all $0\le v\le V$, we arrive at
\begin{eqnarray}
\label{e31} Z_2(e^V,\beta)=\hat Z_2(\beta)e^{(V-V_0)(2\phi(\beta)-\phi(2\beta)+1)}\nonumber\\
\hat Z_2(\beta)=\frac{1}{1-\phi(2\beta)+2\phi(\beta)}
\end{eqnarray}
Let us derive recursive equations to calculate $Z_n,n>1$, defined
as
\begin{eqnarray}
\label{e32}Z_n(\beta,V)\equiv
e^{(V-V_0)(n\phi(\beta)-\phi(n\beta)+n)}e^{nV_0}\hat Z_n(\beta)
\end{eqnarray}
We need to consider all possible splitting $n=m+(n-m)$, with $1\le
m\le n$. Let as assume that two group with $m$  and $n-m$ endpoints
of our hierarchic tree  are separated by the hierarchic distance
$v$, and their trajectories meet at some point $A$. The calculations
similar to those used to derive Eq.(\ref{e14}) give:
\begin{eqnarray}
\label{e33} \hat Z_n(\beta,V)=\sum_{1\le m\le n}
\hat Z_m(\beta,v)\hat Z_{n-m}(\beta,v)\times\nonumber\\
e^{(V-v)(1+\phi(n\beta)-n\phi(\beta))}
\end{eqnarray}
Then the scaling in Eq.(\ref{e32}) yields:
\begin{eqnarray}
\label{e34} \hat Z_n(\beta)=\frac{\sum_{1\le m\le n}\hat
Z_m(\beta)\hat Z_{n-m}(\beta)}{n-1-\phi(n\beta)+n\phi(\beta)}
\end{eqnarray}
One can use Eqs.(\ref{e31}),(\ref{e34}) to calculate any positive
integer moment of the partition function under the condition
\begin{eqnarray}
\label{e35}n\phi(\beta)+n< 1+\phi(n\beta)
\end{eqnarray}
Eq.(\ref{e32}) implies the multifractal behavior with
\begin{eqnarray}
\label{e36}\xi(n)=n\phi(\beta)-\phi(n\beta)+1
\end{eqnarray}

For the corresponding generating function we now obtain an equation
\begin{eqnarray}
\label{e37} (1+\phi(\beta))x\frac{du(x)}{dx}-\phi(\beta
x\frac{d}{dx})u=\nonumber\\
u(x)+u^2(x),\nonumber\\
u(0)=1,u'(0)=1
\end{eqnarray}
For the same model in \cite{sa12}, another equation has been
derived:
\begin{eqnarray}
\label{e38} \frac{\partial G(x,v)}{\partial v}=\phi (\partial
x)G(x,v)+G(x,v)\ln G(x,v)
\end{eqnarray}
Thus two equations must be related.

\subsection{Comparison of Logarithmic 1-d REM and hierarchic model with normal distribution}

In \cite{fy09} the following partition function has been considered:
\begin{eqnarray}
\label{e39} Z=\epsilon^{\hat\beta^2}\int_0^1dye^{\hat \beta
y(x)},\nonumber\\
<y(x)y(x')>=2\ln \frac{|x-x'|}{\epsilon},
\end{eqnarray}
with the parameter $\epsilon$ being an ultraviolet cutoff needed to regularize the model.

The above model shares the multifractal properties with our model for
normal distribution of random energies, with the mapping
\begin{eqnarray}
\label{e40} \epsilon=1/e^{V_0},\nonumber\\
2\hat \beta^2=\beta^2
\end{eqnarray}
Let us compare the moments. Ref.\cite{fy09} gives the following
expression for the moments:
\begin{eqnarray}
\label{e41} \hat
Z_n=\prod_{j=1}^{j=n}\frac{\Gamma(1-(j-1)\gamma)^2\Gamma(1-\gamma
j)}{\Gamma(2-(n+j-2)\gamma)\Gamma(1-\gamma)}
\end{eqnarray}
where $\gamma=\hat \beta^2$.

We have, naturally $\hat Z_1=1$ and further
\begin{eqnarray}
\label{e42}
\hat Z_2=\frac{\Gamma(1-2\gamma)}{\Gamma(2-\gamma)\Gamma(2-2\gamma)}=\frac{1}{(1-\gamma)(1-2\gamma)}\nonumber\\
\hat
Z_3=\frac{\Gamma(1-2\gamma)^3\Gamma(1-3\gamma)}{\Gamma(2-2\gamma)\Gamma(2-3\gamma)\Gamma(2-2\gamma)}\nonumber\\
\hat
Z_4=\frac{\Gamma(1-2\gamma)^3\Gamma(1-3\gamma)^3\Gamma(1-4\gamma)}{\Gamma(1-\gamma)\Gamma(2-3\gamma)\Gamma(2-4\gamma)\Gamma(2-5\gamma)}
\end{eqnarray}
Comparing with the results of the section II-A, we see that second
moment $\hat Z_2$ has different expansion for the small $\beta^2$,
\begin{eqnarray}
\label{e43} \hat Z_2\approx 1+3\gamma,
\end{eqnarray}
More important information contained in moments is however the {\it
smallest} real pole $\gamma_n$ as a function of $\gamma$
(respectively, $b$). Indeed, precisely those poles define the
critical temperatures $T_n=\gamma^{-1/2}_n$ below which the given
moment of the partition function start to diverge. It is easy to see
from Eq. (\ref{e41}) that for one-dimensional model $\gamma=1/n$,
$n=2,3\ldots$ while Eq.(\ref{e42}) gives the smallest pole poles at
$b/2=1/2$ for $Z_2$, $b/2=1/3$ for $Z_3$, $b/2=1/4$ for $Z_4$, etc.
The results for a few lower moments indicate that the two models
share the same sequence of "transition temperatures". Though we are
unable to prove this statement in full generality, the factor
$(1-bn/2)$ appearing in the denominator of the right-hand side of
the recursion Eq.(\ref{e14}) makes the statement very plausible, if
no special cancelations happen. Assuming that correspondence is
correct, the latter property would imply the probability density for
the partition function in both models share the same power law tail:
${\cal P}(Z)\propto Z^{-1-\frac{1}{\beta^2}}$ valid everywhere in
the high-temperature phase $\beta<\beta_c=1$. Such a tail was indeed
proposed as one of the universal features characterizing the class
of models with logarithmic correlations \cite{fy09}.

We obtained analytical expressions for the hierarchic model's
$Z_n$ at positive b, Eq.(14). Having large series of $Z_n$ for
positive $b$, one can try to construct an approximate expression
for the probability distribution and fractional moments.

\section{Conclusion}

The investigation of the statistical mechanics of REM-like models in
finite dimension and their multifractal properties remains an active
field of research, especially after the advances achieved in
\cite{cf02} and \cite{fy09}. In this article we confirmed that the
multifractal properties are shared by the third class of REM-like
models: directed polymers on a hierarchic tree. We calculated the
corresponding moments important for the applications. In particular,
we use those moments to conjecture the power law tail of the
distribution of partition function $Z$ in the high-temperature
phase. Unfortunately it is impossible to derive exact probability
distribution, as has been done in \cite{fy09} for the 1-d
logarithmic REM. Nevertheless, our formulas allow the calculation of
infinite series of the moments for some values of parameters. We
hope that such information could be applied to recover the
probability distribution, a well known problem in probability theory
\cite{ak61}.

Our results  are interesting for the mathematics of
reaction-diffusion equations as well, as we related them to a
certain nonlinear ODE. Moreover, we found the connection  between
the multifractal spectrum and the hierarchic tree model
Eq.(\ref{e36}), while the latter has been mapped to some
reaction-diffusion equations \cite{sa12}, see Eq.(\ref{e38}) in
the present article. Thus the reaction -diffusion equations have
some multiscaling structures. One can connect reaction diffusion
equation and  nonlinear ODE with the interesting versions of
multi-fractal phenomenon, i.e. \cite{fy06}, and try to investigate
them for deriving the multifractal scaling from the reaction
diffusion equation (\ref{e38}). Such a work is currently in the
progress.

The directed walk models on hierarchic tree describes a rather rich
physics, as well as some connections with quantum models in finite
dimension. The relation to 2-d conformal field models is rather well
known \cite{ch96},\cite{ca97}, and the work \cite{sa12} mentions the
relations to quantum disorder problem in finite dimensions
\cite{ga10}. It will be interesting to try to apply the methods of
the current work or our renormalization group (\ref{e38}) to the
finite dimensional quantum disorder problem.

I thank Y. Fyodorov and V. Zacharovas for the discussion, Academia
Sinica for support.


\begin{thebibliography}{99}

\bibitem{de88} B.Derrida, H. Spohn, {\it J. Stat. Phys.}  {\bf
51}, (1988) 817.
\bibitem{de90} J.Cook, B.Derrida, {\it J.Stat.Phys.} {\bf 61}(1990)961.
\bibitem{de91} J.Cook, B.Derrida, {\it J.Stat.Phys.} {\bf 63}(1991)505.
\bibitem{de93} B.Derrida, M.R. Evans, E.R. Speer, {\it Commun. Math.
Phys.} {\bf15} (1993) 221.
\bibitem{de80} B. Derrida, {\it Phys. Rev. Lett.} {\bf 45} (1980)79; {\it Phys. Rev.B.} {\bf 24},2613(1981).
\bibitem{de89} E.Gardner, B. Derrida, J.Phys.{\bf A22} (1989) 1975.
\bibitem{ma97} B. B. Mandelbrot, Fractals and scaling in finance.
Discontinuity, concentration, risk (Springer, New-York, 1997).

\bibitem{ma97a} B. B. Mandelbrot, L. Calvet, and A. Fisher (1997), preprint, Cowles Foundation Discussion paper 1164.
\bibitem{ba06}
J. F. Muzy, E. Bacry, and A. Kozhemyak {\it Phys. Rev. E} {\bf 73} 066114
(2006).
\bibitem{st07}Ken Kiyono, Zbigniew R. Struzik, and Yoshiharu Yamamoto,
{\it Phys. Rev. E} {\bf 76}, 041113 (2007).
\bibitem{tu}
B. Castaing, Y. Gagne, and E. Hopfinger, {\it Physica D} {\bf 46}, 177
(1990).
\bibitem{ch96} C.C. Chamon, C.Mudry, X.G.Wen , {\it Phys. Rev.Lett.} {\bf 77} (1996) 4194.
\bibitem{ca97} H.E.Castillo et al., {\it Phys.Rev. B} {\bf 56}(1997)10668.
\bibitem{ba}R. J. Baxter, {\it Exactly Solvable Models
 in Statistical Mechanics} ( Academic Press, London, 1982)
 \bibitem{sa02} D. Saakian, {\it Phys. Rev E}, {\bf  65},67104(2002).
 \bibitem{sa09} Saakian D.B.,  {\it J. Stat. Mech.} (2009) P07003.
 \bibitem{sa12}D.B. Saakian, {\it Phys.Rev.E} {\bf 85},011109 (2012).


\bibitem{ca01} D. Carpentier, and P. Le Doussal, {\it Phys. Rev. E} {\bf 63}, 026110 (2001).
\bibitem{bu08} Y.V. Fyodorov and J.P. Bouchaud, {\it J. Phys. A: Math.Gen.} ,{\bf 41}, 372001( 2008).

\bibitem{fy09} Y. V. Fyodorov, P. Le
Doussal, and A. Rosso, {\it J. Stat. Mech.} P10005
(2009).
\bibitem{se44}A. Selberg,
{\it Norsk. Mat. Tidsskr.} {\bf 24} (1944), 71-78.
\bibitem{se08}P.J. Forrester, S.O. Warnaar, {\it Bull. Am. Math. Soc.} {\bf 45}
489 (2008).
 \bibitem{Ost} D. Ostrovsky {\it Rev. Math. Phys.} {\bf 23} 127 (2011)

\bibitem{ba00}Muzy J-F, Delour J and Bacry E  {\it Eur. Phys. J. B}
17 537(2000).
\bibitem{ba01}
E. Bacry, J. Delour, and J. F. Muzy {\it Phys. Rev. E} {\bf 64} 026103
(2001).
\bibitem{sa12a}D.B. Saakian, Phys. Rev. E (2012).
\bibitem{cf02} L. Calvet and A. Fisher, {\it Rev. Econ. Stat.} 84, 381 (2002).
\bibitem{fy09a} Y.V. Fyodorov, {\it J. Stat. Mech.} P07022  (2009).
 \bibitem{fy10}Y.V. Fyodorov,{\it  Physica A},{\bf 389}, 4229 (2010).
\bibitem{je86} T.C. Halsey, M.H. Jensen, L.P. Kadanov, I. Procaria and B.
Shraiman, {\it Phys. Rev. A} 33, 1141 (1986).
 \bibitem{pa87} G. Paladin
and A. Vulpiani, {\it Phys. Rep.} 156, 147 (1987).
\bibitem{ak61}N.I. Akhiezer, The classical problem of moments and
some related prpblems of analysis, in Russian, Moscow, (1961).
\bibitem{ba06}
E.Bacry, A. Kozhemyak, J.F. Muzy {\it J. Econ. Dyn. $\&$
Control}, {\bf 32}, 156 (2008).

\bibitem{fr}E. L. Ince, Ordinary Differential Equations, Dover, New York, 1956.
\bibitem{hw10}H.H. Chern, H.K. Hwang,  C.Martinez,    arXiv: 1002. 3859 v1 [math.pr] Psi-series method in random trees and moments of
high orders.
\bibitem{ga10}C.Monthus,C.Garel, {\it J. Stat. Mech.},(2010)P06014.
\bibitem{fy06} A.D. Mirlin, Y.V. Fyodorov, A. Mildenberger and F. Evers {\it Phys. Rev. Lett} {\bf 97}, 046803
(2006).

\end{thebibliography}
\end{document}